\documentclass[12pt,onecolumn,journal]{IEEEtran}

\usepackage{mathrsfs}
\usepackage{txfonts}
\usepackage [dvips] {graphicx}

\ifCLASSINFOpdf

\else

\fi

\makeatletter
\newenvironment{tablehere}
{\def\@captype{table}} {}

 \makeatother

\begin{document}

\title{Robust Capon Beamforming via Shaping Beam Pattern}

\author{Yipeng~Liu

\thanks{Yipeng Liu is with ESAT-SCD/iMinds Future health department, Dept. of Electrical Engineering, KU Leuven, Kasteelpark Arenberg 10, box 2446, 3001 Heverlee, Belgium. Most of this work was done when he was with Dept. of Electronic Engineering, University of Electronic Science and Technology of China (UESTC), Chengdu, 611731, China.
e-mail: (yipeng.liu@esat.kuleuven.be; dr.yipengliu@gmail.com).}

}

\markboth{Yipeng Liu: Robust Capon Beamforming via Shaping Beam Pattern}%
{Shell \MakeLowercase{\textit{et al.}}: Bare Demo of IEEEtran.cls for Journals}

\maketitle

\begin{abstract}

High sidelobe level and direction of arrival (DOA) estimation sensitivity are two major disadvantages of the Capon
beamforming. To deal with these problems, this paper gives an overview of a series of robust Capon
beamforming methods via shaping beam pattern, including sparse Capon beamforming, weighted sparse Capon beamforming, mixed norm based Capon beamforming, total variation minimization based Capon beamforming, mainlobe-to-sidelobe power ratio maximization based Capon beamforming. With these additional structure-inducing constraints, the sidelobe is suppressed, and the robustness against DOA mismatch is improved too. Simulations show that the obtained
beamformers outperform the standard Capon beamformer.

\end{abstract}


\IEEEpeerreviewmaketitle

\section{Introduction}

Beamforming is a conventional technique for signal's emission and reception in  some certain directions in sensor array system \cite{li_rab} \cite{stoica_Spectral Analysis} \cite{vantrees_Optimum Array Processing}. To selectively receive and transmit in space, beam pattern can be synthesized either adaptively or deterministically. Comparing with the omnidirectional sensor system, beamforming system exploits the signal's spatial dimension to enhance the signal quality. It has a wide range of applications in the field of radar, sonar, acoustics, astronomy, seismology, communications, and medical imaging.

The non-adaptive (data-independent) beamforming consists of the delay-and-sum method and a variety of weighting based sidelobe control methods; while adaptive (data-dependent) beamforming solves a optimization problem with a data-driven performance function to get a set of array weighting vector. The data-independent beamformer uses a set of pre-defined weights to linearly combine the transmitted/received signals in different sensors, which only uses the information about spatial position of the sensor and signal-of-interest (SOI). Generally data-dependent beamforming will further exploit some characteristics of the transmitted/received signal, to suppress the interference and noise in the non-interested directions.

The Capon beamformer is one of the most popular adaptive beamforming systems. It minimizes the array output power while subjecting to the linear constraint that the SOI does not suffer from any distortion by adaptive selection of the weighting vector. The Capon beamformer has better resolution and much better interference rejection capability than the data-independent beamformer. However, its high sidelobe level and the SOI steering vector uncertainty due to differences between the assumed signal arrival angle and the true arrival angle would seriously degenerate the performance in the presence of environment noise and interferences \cite{wax_mvb} \cite{cox_mismatch}.

With a spherical uncertainty set being introduced, doubly constrained robust (DCR) Capon beamformer is obtained with the increased robustness against direction of arrival (DOA) mismatch \cite{li_dcrcb}. The diagonal loading method is one of the most popular ways to deal with the uncertainty in steering vector in robust Capon beamforming (RCB) \cite{li_rab} \cite{li_diagonal loading}. But its main drawback is that the diagonal loading factor is not convenient to determine. Eigen-space-based beamforming approach \cite{Chang_eigenspace-based beamformer}, which can only be used to the point signal source and high signal-to-noise-ratio (SNR) cases, uses the projection of the presumed signal steering vector onto the sample signal-plus-interference subspace instead of the presumed signal steering vector. Many works have been done to enhance the robustness against the SOI steering vector uncertainty, to accelerate the convergence, etc \cite{li_rab}.

In this paper, different from the previous RCB, a class of new methods to obtain the robustness by shaping the beam pattern are summarized. In \cite{zhang_sparse beamforming}, sparse constraint is used to encourage sparse distribution of the array gains. In \cite{liu_weighted sparse beamforming}, weighting is used to make the distribution sparser. They are the first two papers that use the beam pattern shaping constraints. But the structure information of the beam pattern is not fully exploited. In order to better match to the ideal beam pattern, several other constraints are proposed, including the mixed norm based constraint \cite{liu_mixed norm beamforming}, total variation minimization (TVM) constraint \cite{Chambolle_tvm} \cite{liu_tvm beamforming}, and the mainlobe to sidelobe power ratio maximization constraint \cite{liu_mspr beamforming}. These beam pattern shaping constraints use different ways to encourage dense distribution of the array gains in the mainlobe and sparse distribution of the array gain in the sidelobe. i.e., nearly all the large gains are accumulated in the vicinity of the mainlobe, while the smaller ones are in the sidelobes. Thereby, when there is uncertainty in the estimated direction of SOI, the array gain in the real direction of SOI, which is often near the estimated one,  can be large enough to keep out large SOI power loss. At the same time, the interferences and noise is largely blocked, because nearly all the array gains in the sidelobes are much smaller. Numerical experiments show that the beam pattern shaping constraint based robust Capon beamforming outperforms standard one. Different kinds of beam pattern shaping constraints have their own advantages.

\section{Signal Model}

\begin{figure}[!h]
 \centering
 \includegraphics[angle= 0, scale = 0.27]{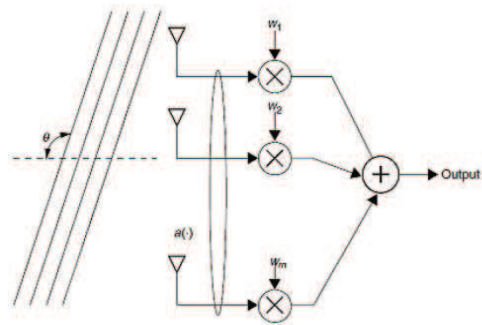}
 \caption{Array structure of the beamformer.}
 \label{figure1}
\end{figure}

As it shows in Fig. \ref{figure1}, the signal impinging into a uniform linear array (ULA) with \emph{M} antennas can be represented by an \emph{M}-by-1 vector \cite{li_rab}:
\begin{equation}
\label{eq2_1_received_signal} {\bf{x}}(k) = s(k){\bf{a}}({\theta _0}) + \sum\limits_{j = 1}^J {{\beta _j}(k){\bf{a}}({\theta _j})}  + {\bf{n}}(k)
\end{equation}
where \emph{k} is the index of snapshot, \emph{J} is the number of interferences, \emph{s}(\emph{k}) and $ {{\beta _j}(k)} $, \emph{j} = 1, ... , \emph{J}, are the amplitudes of the SOI and interferences at \emph{k}, respectively, $ {{\theta _j}}$ (for \emph{l} = 0, 1, ... , \emph{J}) are the values of DOA of the SOI and interferences. \textbf{n}(\emph{k}) is the additive white Gaussian noise (AWGN) vector at time instant \emph{k}. $ {\bf{a}}({\theta _j}) $, \emph{j} = 1, ... , \emph{J} is the array steering vector, whose \emph{m}-th element is
\begin{equation}
\label{eq2_2_steering vector} {\left[ {{\bf{a}}({\theta _j})} \right]_m} = \exp \left( {i(m - 1)\frac{{2\pi d}}{\lambda }\sin {\theta _j}} \right)
\end{equation}
where $ i = \sqrt { - 1} $, \emph{d} is the distance between two adjacent sensors and  $ \lambda $ is the wavelength of the SOI. The output signal of a beamformer at the time instant \emph{k} can be formulated as:
\begin{equation}
\label{eq2_2_output of beamformer}
y(k) = {{\bf{w}}^H}{\bf{x}}(k) = s(k){{\bf{w}}^H}{\bf{a}}({\theta _0}) + \sum\limits_{j = 1}^J {{\beta _j}(k){{\bf{w}}^H}{\bf{a}}({\theta _j})}  + {{\bf{w}}^H}{\bf{n}}(k)
\end{equation}
where \textbf{w} is the \emph{M}-by-1 complex-valued weighting coefficients of beamformer.

\section{Classical Capon beamforming}

In classical adaptive beamformer, it enforces array gain in the estimated direction to be constant $ {{\bf{w}}^H}{\bf{a}}({\theta _0}) \approx 1 $ and minimize the array output power of interference and noise $ {{\bf{w}}^H}{\bf{n}}(k) $ to make the array output is approximately equal to the SOI. Here the array response gain in the direction $ {\theta _0} $ is $ {{\bf{w}}^H}{\bf{a}}({\theta _0}) $, and the noise and interference gains in the array output is $ {{\bf{w}}^H}{{\bf{R}}_n}{\bf{w}} $, wherein $ {{\bf{R}}_n} = {\rm E}\left( {{\bf{n}}{{\bf{n}}^H}} \right) $, and $ {\rm E}$ denotes the expectation operator. Assuming $ {\bf{a}}({\theta _0}) $ and $ {{\bf{R}}_n} $ are known in advance, the optimization model to produce the optimal array weighting vector is:
\begin{equation}
\label{eq3_1_optimal_beamformer}
\begin{array}{c}
 \mathop {\min }\limits_{\bf{w}} \left( {{{\bf{w}}^H}{{\bf{R}}_n}{\bf{w}}} \right),~~
 {\rm{ s}}{\rm{.~t}}{\rm{.~~  }}{{\bf{w}}^H}{\bf{a}}({\theta _0}) = 1 \\
 \end{array}
\end{equation}
where $ {{\bf{R}}_n} $ and  $ {\theta _0}$ are assumed to be known exactly in advance. However, it is not always in this case. The Capon beamformer is proposed to replace the noise and interference power $ {{\bf{R}}_n} $  with the covariance of the received signals $ {{\bf{R}}_x} $. The estimated covariance of the received signals $ {{\bf{R}}_x} \in {\textbf{C}^{M \times M}} $ can be obtained by processing multiple snapshots received by multiple sensors, i.e.:
\begin{equation}
\label{eq3_2_covariance_matrix}
{{\bf{R}}_x} = \frac{1}{M}\sum\limits_{j = k - M + 1}^k {{\bf{x}}(j){{\bf{x}}^H}(j)}
\end{equation}
Thus, the Capon beamforming can be formulated as:
\begin{equation}
\label{eq3_3_mvdr}
{{\bf{w}}_{MVDR}} = \mathop {\arg \min }\limits_{\bf{w}} \left( {{{\bf{w}}^H}{{\bf{R}}_x}{\bf{w}}} \right),~~{\rm{  s}}{\rm{.~t}}{\rm{.~~  }}{{\bf{w}}^H}{\bf{a}}({\alpha _0}) = 1
\end{equation}
where $ {{\bf{w}}^H}{{\bf{R}}_x}{\bf{w}} $ is the minimum covariance constraint, $ {{\bf{w}}^H}{\bf{a}}({\alpha _0}) = 1 $ is the distortionless constraint of the SOI. That why the Capon beamforming is called minimum variance distortionless response (MVDR) beamforming too; $ {\alpha _0} $ is the estimated direction of SOI, which cannot be exactly equal to the real direction of SOI $ {\theta _0} $ because of estimation error. Solving the optimization model of MVDR beamforming can give a close form solution for the optimal weighting vector as:
\begin{equation}
\label{eq3_3_mvdr_weighting}
{{\bf{w}}_{MVDR}} = \frac{{{\bf{R}}_x^{ - 1}{\bf{a}}({\alpha _0})}}{{{{\bf{a}}^H}({\alpha _0}){\bf{R}}_x^{ - 1}{\bf{a}}({\alpha _0})}}
\end{equation}
The difference between (\ref{eq3_1_optimal_beamformer}) and (\ref{eq3_3_mvdr}) is that the MVDR beamforming minimizes the power of SOI, interference and noise. and meanwhile its distortionless constraint can keep there is little loss in the direction of SOI, which guarantees the array gain in the direction of SOI is not degenerated.

\section{Capon beamforming with shaped beam pattern}

\subsection{Sparse Capon beamforming}

In \cite{zhang_sparse beamforming}, a sparse constraint is incorporated into the MVDR beamformer to enforce sparse distribution in the whole beam pattern. It can reduce the number of nonzero array gains and suppress the sidelobe level of the classical MVDR beamformer. The sparse constraint (SC) based improved MVDR beamformer can be formulated as:
\begin{equation}
\label{eq4_1_Lp sparse_beamforming}
{{\bf{w}}_{SC}} = \mathop {\arg \min }\limits_{\bf{w}} \left( {{{\bf{w}}^H}{{\bf{R}}_x}{\bf{w}} + {\gamma _1}\left\| {{{\bf{w}}^H}{\bf{A}}} \right\|_p^p} \right),~~{\rm{  s}}{\rm{.~t}}{\rm{. ~~ }}{{\bf{w}}^H}{\bf{a}}({\theta _0}) = 1
\end{equation}
where the non-negative scalar $ {{\gamma _1}} $ is the parameter that makes balance of the minimum variance constraint and the sparse constraint, $ {\left\| {\bf{x}} \right\|_p} = {\left( {\sum\nolimits_i {{{\left| {{x_i}} \right|}^p}} } \right)^{1/p}} $ is the $ Lp $ norm of a vector \textbf{x}, the \emph{M}-by-\emph{N} \textbf{A} is the array manifold with $ {\alpha _n} $s ( \emph{n} = 1, 2, ... , \emph{N}) being the sampled angles ranging from $ - {90^ \circ }$ to $  {90^ \circ }$, which covers all the \emph{N} steering vectors for signals impinging from all possible angles, i.e.
\begin{equation}
\label{eq4_2 array manifold}
{\bf{A}} = \left[ {\begin{array}{*{20}{c}}
   1 &  \cdots  & 1  \\
   {\exp \left( {j{\varphi _1}} \right)} &  \cdots  & {\exp \left( {j{\varphi _N}} \right)}  \\
    \vdots  &  \ddots  &  \vdots   \\
   {\exp \left( {j\left( {M - 1} \right){\varphi _1}} \right)} &  \cdots  & {\exp \left( {j\left( {M - 1} \right){\varphi _N}} \right)}  \\
\end{array}} \right]
\end{equation}

\begin{equation}
\label{eq4_3 array manifold_1}
{\varphi _n} = \frac{{2\pi d}}{\lambda }\sin {\alpha _n},~ for~~ \emph{n}=1,2,...,\emph{N}
\end{equation}
 Different from previous sparse constraint in synthesis form, i.e. the sparsity is obtained by decomposing the estimated variables into a synthesis dictionary and a sparse vector, the used sparsity constraint here is in analysis, i.e. the enforced sparsity in the \emph{Lp} norm constraint is obtained by multiplying the estimated variables with a synthesis dictionary. Recently this kind of sparsity in analysis form is called cosparsity too \cite{nam_cosparsity}.

Considering that when $ 0 \le p < 1 $, the minimization of the $ Lp $ norm is not convex, $ p = 1 $ is chosen to make the optimization model for sparse beamformer be convex, i.e.:

\begin{equation}
\label{eq4_4 L1 sparse beamformer}
\begin{array}{c}
 {{\bf{w}}_{SC}} = \mathop {\arg \min }\limits_{\bf{w}} \left( {{{\bf{w}}^H}{{\bf{R}}_x}{\bf{w}} + {\gamma _1}\left\| {{{\bf{w}}^H}{\bf{A}}} \right\|_1^{}} \right) ,~~
 {\rm{  s}}{\rm{.~t}}{\rm{.~~  }}{{\bf{w}}^H}{\bf{a}}({\alpha _0}) = 1 \\
 \end{array}
\end{equation}
The sparse Capon beamforming (\ref{eq4_4 L1 sparse beamformer}) is a convex programming, and can be solved efficiently \cite{boyd_cvx}.

\subsection{Weighted sparse Capon beamforming}

We consider the \emph{M}-by-1 vector, \textbf{x}(\emph{k}), as a snapshot of the received signal at time instant \emph{k}. If we collect the snapshots of \emph{K} ($ K \ge 1 $) different time instants in a matrix, then we can have an \emph{M}-by-\emph{K} data matrix as
\begin{equation}
\label{eq5_1 data matrix}
{\bf{X}} = {\rm{ }}\left[ {{\bf{x}}\left( {\rm{1}} \right),{\bf{x}}\left( {\rm{2}} \right), \ldots ,{\bf{x}}\left( K \right)} \right]
\end{equation}
It has been shown that the cross-correlation of the steering matrix \textbf{A} and the received data matrix \textbf{X} coarsely represents the \emph{a posteriori} spatial distribution of interfering signals \cite{haykin adaptive fiter}. We use this property to define a weighted sparse constraint for further suppressing sidelobe level of the beam pattern. As a result, the weighting vector of the beamformer with a weighted sparse constraint is given by
\begin{equation}
\label{eq5_2 weighted sparse beamformer}
{{\bf{w}}_{WSC}} = \mathop {\arg \min }\limits_{\bf{w}} \left( {{{\bf{w}}^H}{{\bf{R}}_x}{\bf{w}} + {\gamma _2}\left\| {{{\bf{w}}^H}{\bf{AQ}}} \right\|_p^p} \right),~{\rm{  s}}{\rm{.~t}}{\rm{.~~  }}{{\bf{w}}^H}{\bf{a}}({\theta _0}) = 1
\end{equation}
where the \emph{N}-by-\emph{N} matrix \textbf{Q} = diag[$ SNM\left( {{{\bf{A}}^H}{\bf{X}}} \right) $] serves as a weighting matrix, and $ SNM\left( {{{\bf{A}}^H}{\bf{X}}} \right) $ is an \emph{N}-by-1 vector containing elements the squared normalized mean value of each row of the \emph{N}-by-\emph{K} matrix $ {{{\bf{A}}^H}{\bf{X}}} $ \cite{liu_weighted sparse beamforming} \cite{haykin adaptive fiter}, and $ {\gamma _2} $ is the weighting factor balancing the minimum variance constraint and the beam pattern shaping constraint. Comparing (\ref{eq5_2 weighted sparse beamformer}) with (\ref{eq4_4 L1 sparse beamformer}), we can see that the matrix \textbf{Q} in (\ref{eq5_2 weighted sparse beamformer}) provides additional weighting on the sparse constraint, in accordance with the DOA distribution of interfering signals. More specifically, the larger the probability of interference arriving in a certain direction, the larger the weight applied on the sparse constraint in the corresponding direction.

The optimal weighting vector indicated by (\ref{eq5_2 weighted sparse beamformer}) can be found by using an adaptive iteration algorithm \cite{liu_weighted sparse beamforming}. When \emph{p} = 1, a series of algorithms for convex programming, can be used to solve (\ref{eq5_2 weighted sparse beamformer}) efficiently \cite{boyd_cvx}. We also observe that (\ref{eq4_4 L1 sparse beamformer}) can be considered as a special case of (\ref{eq5_2 weighted sparse beamformer}), in terms that (\ref{eq5_2 weighted sparse beamformer}) reduces to (\ref{eq4_4 L1 sparse beamformer}) when \textbf{Q} = \textbf{I}, corresponding to the case of equal weighting in every direction.

\subsection{Mixed norm based Capon beamforming}

In (\ref{eq4_4 L1 sparse beamformer}), the added constraint encourages sparse distribution in all the possible values of DOA from $ - {90^ \circ } $ to $ {90^ \circ } $, no matter whether the array gains are in the mainlobe or the sidelobe. However, the array gains are not in this kind of conventional sparse distribution. The array gains are densely distributed in the mainlobe and sparsely distributed in the sidelobe. To exploit this more detailed structural sparsity information to improve the performance, a mixed norm constraint with two kinds of norms on mainlobe and sidelobe respectively can be added to the Capon beamformer. The new beamformer can be formulated as \cite{liu_mixed norm beamforming}:

\begin{equation}
\label{eq6_1 mixed norm based Capon beamforming}
\begin{array}{c}
 {{\bf{w}}_{MNC}} = \mathop {\arg \min }\limits_{\bf{w}} \left[ {{{\bf{w}}^H}{{\bf{R}}_x}{\bf{w}} + {\gamma _3}\left( {\left\| {{{\bf{w}}^H}{{\bf{A}}_M}} \right\|_\infty ^{} + \left\| {{{\bf{w}}^H}{{\bf{A}}_S}} \right\|_1^{}} \right)} \right] ,~~
 {\rm{  s}}{\rm{.~t}}{\rm{. ~~ }}{{\bf{w}}^H}{\bf{a}}({\theta _0}) = 1 \\
 \end{array}
\end{equation}
where $ {\gamma _3} $ is the weighting parameter, $ {{\bf{A}}_M} $ and $ {{\bf{A}}_S} $ contains the vectors from the steering matrix \textbf{A}, $ {{\bf{A}}_M} $ is composed of 2\emph{b} + 1 steering vectors corresponding to the mainlobe; and $ {{\bf{A}}_S} $ is constituted with the steering vectors in \textbf{A} corresponding to the sidelobe. The product $ {{\bf{w}}^H}{{\bf{A}}_M} $ indicates array gains of the mainlobe in the beam pattern, and $ {{\bf{w}}^H}{{\bf{A}}_S}$ indicates array gains of the sidelobe. \emph{b} is an integer in represention of the bound between the mainlobe and the sidelobe. T minimization of $ \left\| {{{\bf{w}}^H}{{\bf{A}}_S}} \right\|_1^{} $  enforces the sparse distribution of the array gains in the sidelobe, and the minimization of $ \left\| {{{\bf{w}}^H}{{\bf{A}}_M}} \right\|_\infty ^{} $ results in dense distribution of the array gains in the mainlobe.

\begin{figure}[!h]
 \centering
 \includegraphics[angle= 270, scale = 0.27]{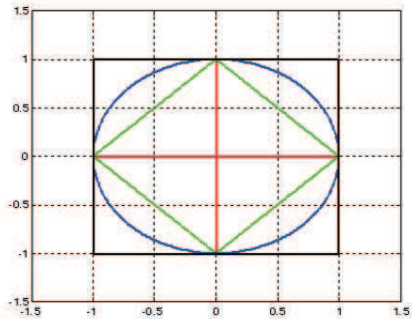}
 \caption{The geometric illustration of the $ Lp $ norm minimization when $ p = 0,1,2,\infty  $.}
 \label{figure2}
\end{figure}

To illustrate why different kinds of norms encourage different kinds of distributions, a simple 2-dimensional geometry illustration is given in Fig. \ref{figure2}. The minimization of the L1 norm, the L2 norm and the $ L\infty $ norm of a two dimensional vector can be represented as a smallest rhombus, a smallest circular and a smallest exact square, respectively. The distortionless constraint is a fixed line in the plane. The optimal solution for the minimization of the norm $ \left\| {{{\bf{w}}^H}{\bf{A}}} \right\|_p^{} $, (\emph{p} = 1, 2, $ \infty $) subjecting to the distortionless constraint $ {{\bf{w}}^H}{\bf{a}}({{\bf{\theta }}_0}) = 1 $ would be the tangent point of the line and the curve (rhombus, circular, square). Therefore, the L1 norm minimization results in two entries of the solution which have quite different absolute values with high probability; and the $ L\infty $ norm minimization gives birth to two entries of the solution which have considerable similar absolute values with high probability.

With this mixed norm constraint added, the obtained beam pattern would be sharped like this: most of the significant array gains are located in the mainlobe area; and the rest trivial array gains are in the sidelobe area. Since most of entries in the mainlobe are non-trivial, the array gains in the mismatched angle are also significant, which avoids seriously degenerate the performance. In addition, Since most of entries in the sidelobe are trivial, nearly all of the array gains in the angles of interferences and noise are very small, which reduce the power of interferences and background noise in the output signal.

Here we call (\ref{eq6_1 mixed norm based Capon beamforming}) the mixed norm (MN) based Capon (MNC) beamformer, and its solution can be given by some efficient algorithms \cite{boyd_cvx}.

\subsection{Total variation based Capon beamforming}

The sparse constraint encourages sparse distribution for all the array gains $ {{\bf{w}}^H}{\bf{A}} $ for all the possible values of DOA from $ - {90^ \circ }$ to $ {90^ \circ }$. However, the array gains in the mainlobe are not sparse, but dense as a solid block. To improve the performance with a more suitable constraint on the beam pattern, the L1 norm minimization based sparse constraint is only added on the sidelobe, and a TVM restricts for the entire beam pattern \cite{Chambolle_tvm} \cite{liu_tvm beamforming}. The TVM and sparse constraint based Capon beamformer can be formulated as \cite{liu_tvm beamforming}:
\begin{equation}
\label{eq7_1 tv based Capon beamforming}
\begin{array}{c}
 {{\bf{w}}_{TVMS}} = \mathop {\arg \min }\limits_{\bf{w}} \left( {{{\bf{w}}^H}{{\bf{R}}_x}{\bf{w}} + {\gamma _4}\left( {\sum\limits_{i = 1}^I {{{\left\| {{{\bf{D}}_i}{{\left( {{{\bf{w}}^H}{\bf{A}}} \right)}^T}} \right\|}_2}}  + {{\left\| {{{\bf{w}}^H}{{\bf{A}}_S}} \right\|}_1}} \right)} \right),{\rm{ }} \\
 {\rm{ s}}{\rm{.~t}}{\rm{. ~~ }}{{\bf{w}}^H}{\bf{a}}({\alpha _0}) = 1 \\
 \end{array}
\end{equation}
where

\begin{equation}
\label{eq7_2 difference matrix}
{{\bf{D}}_i} = \left[ {\begin{array}{*{20}{c}}
   {{{\bf{D}}_{i,F}}}  \\
   {{{\bf{D}}_{i,B}}}  \\
\end{array}} \right]
\end{equation}

$ {{\bf{D}}_{i,F}} $ and $ {{\bf{D}}_{i,B}} $ are the \emph{i}-th order forward and backward differential matrices. \emph{I} is the total number of differential matrix $ {{\bf{D}}_{i}} $; $ {{\bf{A}}_{\rm{S}}} $ is constituted with the sidelobe steering vectors in \textbf{A}. The product $ {{\bf{w}}^{\rm{H}}}{{\bf{A}}_{\rm{S}}} $ indicates array gains of the sidelobe. $ {\gamma _4}$ is the weighting factor controlling the TVM constraint and the sparse constraint. Since the objective function of the proposed beamformer (\ref{eq7_1 tv based Capon beamforming}) is convex; the optimal $ {{\bf{w}}_{{\rm{TVMS}}}} $ can also be solved out by convex programming software \cite{boyd_cvx}.

In the beam pattern shaping constraint $ \sum\limits_{i = 1}^I {{{\left\| {{{\bf{D}}_i}{{\left( {{{\bf{w}}^H}{\bf{A}}} \right)}^T}} \right\|}_2}}  + {\left\| {{{\bf{w}}^H}{{\bf{A}}_S}} \right\|_1} $, the first term is the TVM which discourages the large fluctuation in the beam pattern. It results in the high array gains accumulated in the mainlobe and the small trivial ones gathered in the sidelobe. In addition, the sparse beam pattern constraint is modified to be the second term to further suppress the sidelobe level. As the new constraint in (\ref{eq7_1 tv based Capon beamforming}) fits the desired beam pattern better, the performance would be enhanced.

\subsection{Mainlobe-to-sidelobe power ratio maximization based Capon beamforming}

In the perspective of the beam pattern, it is observed from the Capon beamformer that there is only an explicit constraint on the desired DOA, i.e. $ {{\bf{w}}^H}{\bf{a}}({\theta _0}) = {\rm{1}} $, while no constraint is put onto the interference and background noise. To repair this drawback, we propose the following cost function with a regularization term, which forces maximization of the MSPR \cite{liu_mspr beamforming}:

\begin{equation}
\label{eq8_1 mspr1}
\begin{array}{c}
 {{\bf{w}}_{MSPR}} = \mathop {\arg \min }\limits_{\bf{w}} \left\{ {{{\bf{w}}^H}{{\bf{R}}_x}{\bf{w}} + {\gamma _5}\frac{{\left\| {{{\bf{w}}^H}{{\bf{A}}_S}} \right\|_2^2}}{{\left\| {{{\bf{w}}^H}{{\bf{A}}_M}} \right\|_2^2}}} \right\} \\
 {\rm{  s}}{\rm{.~t}}{\rm{.~~  }}{{\bf{w}}^H}{\bf{a}}({\theta _0}) = 1 \\
 \end{array}
\end{equation}
where $ \gamma _5 $ is the weighting factor balancing the minimum variance constraint and the MSPR maximization constraint.

Physically $ {{\bf{w}}^H}{\bf{a}}(\theta )$ is the array gain in the signal direction $ \theta $. The product $ {{\bf{w}}^H}{{\bf{A}}_S} $ indicates array gains of the sidelobe; and the product $ {{\bf{w}}^H}{{\bf{A}}_M} $ indicates array gains of the mainlobe. With the term $ {{\left\| {{{\bf{w}}^H}{{\bf{A}}_S}} \right\|_2^2} \mathord{\left/
 {\vphantom {{\left\| {{{\bf{w}}^H}{{\bf{A}}_S}} \right\|_2^2} {\left\| {{{\bf{w}}^H}{{\bf{A}}_M}} \right\|_2^2}}} \right.
 \kern-\nulldelimiterspace} {\left\| {{{\bf{w}}^H}{{\bf{A}}_M}} \right\|_2^2}} $  minimized, the MSPR based Capon beamformer (\ref{eq8_2 mspr2}) can perform sidelobe minimization and mainlobe maximization simultaneously. But unfortunately it is not convex. To make it convex, we relax the MSPR constraint and obtain a new beamformer as

\begin{equation}
\label{eq8_2 mspr2}
\begin{array}{c}
 {{\bf{w}}_{RMSPR}} = \mathop {\arg \min }\limits_{\bf{w}} \left\{ {{{\bf{w}}^H}{{\bf{R}}_x}{\bf{w}} + {\gamma _5}\left[ {{{\left( {\left\| {{{\bf{w}}^H}{{\bf{A}}_M}} \right\|_2^2 - 1} \right)}^2} + \left\| {{{\bf{w}}^H}{{\bf{A}}_S}} \right\|_2^2} \right]} \right\} \\
 {\rm{  s}}{\rm{.~t}}{\rm{.~~  }}{{\bf{w}}^H}{\bf{a}}({\theta _0}) = 1 \\
 \end{array}
\end{equation}

The splitting of the matrix \textbf{A} into $ {{\bf{A}}_M} $ and $ {{\bf{A}}_M} $ helps. The newly added relaxed MSPR (RMSPR) term $ {\left( {\left\| {{{\bf{w}}^H}{{\bf{A}}_M}} \right\|_2^2 - 1} \right)^2} + \left\| {{{\bf{w}}^H}{{\bf{A}}_S}} \right\|_2^2 $  is convex. It is minimized to minimize the sidelobe power $ \left\| {{{\bf{w}}^H}{{\bf{A}}_S}} \right\|_2^2 $ and the approximation error $ {\left( {\left\| {{{\bf{w}}^H}{{\bf{A}}_M}} \right\|_2^2 - 1} \right)^2}$. It is one kind of way to reshaping the Capon beam pattern with constraint on both mainlobe and sidelobe. The mainlobe power and sidelobe power are constrained separately in the optimization model (\ref{eq8_2 mspr2}). It is no doubt that the minimization of $ \left\| {{{\bf{w}}^H}{{\bf{A}}_S}} \right\|_2^2 $ gives smaller power of sidelobe. The minimization $ {\left( {\left\| {{{\bf{w}}^H}{{\bf{A}}_M}} \right\|_2^2 - 1} \right)^2} $  makes the power of mainlobe be a constant approximately. i.e. for $ {\left( {\left\| {{{\bf{w}}^H}{{\bf{A}}_M}} \right\|_2^2 - 1} \right)^2} < \xi $, if $ \xi $ is very small, the power of mainlobe $ \left\| {{{\bf{w}}^H}{{\bf{A}}_M}} \right\|_2^2 $ can be approximately constant. Specially if $ \xi $ = 0 , the mainlobe power is strictly constant. Combining these two constraints for beam pattern shaping, it forces the power of mainlobe $ \left\| {{{\bf{w}}^H}{{\bf{A}}_M}} \right\|_2^2 $ to be an approximately constant while making the power in sidelobe  $ \left\| {{{\bf{w}}^H}{{\bf{A}}_S}} \right\|_2^2 $ as small as possible. With the power in mainlobe being a constant and the power in sidelobe being minimized, the MSPR can be maximized in a relaxed way. Thus, the relaxed form of the maximization of MSPR $ {{\left\| {{{\bf{w}}^H}{{\bf{A}}_S}} \right\|_2^2} \mathord{\left/
 {\vphantom {{\left\| {{{\bf{w}}^H}{{\bf{A}}_S}} \right\|_2^2} {\left\| {{{\bf{w}}^H}{{\bf{A}}_M}} \right\|_2^2}}} \right.
 \kern-\nulldelimiterspace} {\left\| {{{\bf{w}}^H}{{\bf{A}}_M}} \right\|_2^2}} $  would be achieved.

\section{Numerical experiments}

Numerical experiments are employed to demonstrate the performance of different beam pattern shaping constraint based Capon beamforming methods and the standard Capon beamforming. a ULA with 8 half-wavelength spaced antennas is considered. The AWGN at each sensor is assumed spatially uncorrelated. The DOA of the SOI is set to be $ {0^ \circ } $, and the DOAs of three interfering signals are set to be  $  - {30^ \circ
} $, $  {30^ \circ } $, and $  {70^ \circ } $, respectively. The SNR is set to be 10 dB, and the interference to noise ratios (INRs) are assumed to be 20 dB, 20 dB, and 40 dB in $  - {30^ \circ
} $, $  {30^ \circ } $, and $  {70^ \circ } $, respectively. 100 snapshots are used for each simulation. The matrix \textbf{A} consists of all steering vectors ranging in [$  - {90^ \circ } $, $  {90^ \circ } $] with the sampling interval of $  {1^ \circ } $. Without loss of generality, \emph{p} is set to be 1; \emph{b} is set to be 15; and $ {\gamma _i} $, \emph{i}=1,2,...,5 are all set to get the optimal performance.

To quantify the performance evaluation of different beamformers, the signal to interference noise ratio (SINR) of beamformer's output can be defined as
\begin{equation}
\label{eq9_1_SINR} SINR = \frac{{\sigma
_s^2{{\bf{w}}^H}{\bf{a}}({\theta _0}){{\bf{a}}^H}({\theta
_0}){\bf{w}}}}{{{{\bf{w}}^H}\left( {\sum\limits_{j = 1}^J {\sigma
_j^2{\bf{a}}({\theta _j}){{\bf{a}}^H}({\theta _j})}  + {\bf{Q}}}
\right){\bf{w}}}}
\end{equation}
where $ \sigma _s^2 $ and $ \sigma _j^2 $  are the variances of the
SOI and the \emph{j}-th interference, and \textbf{Q} is a diagonal matrix whose
  diagonal elements are the variances of the noise.

\begin{figure}[!h]
 \centering
 \includegraphics[scale = 0.47]{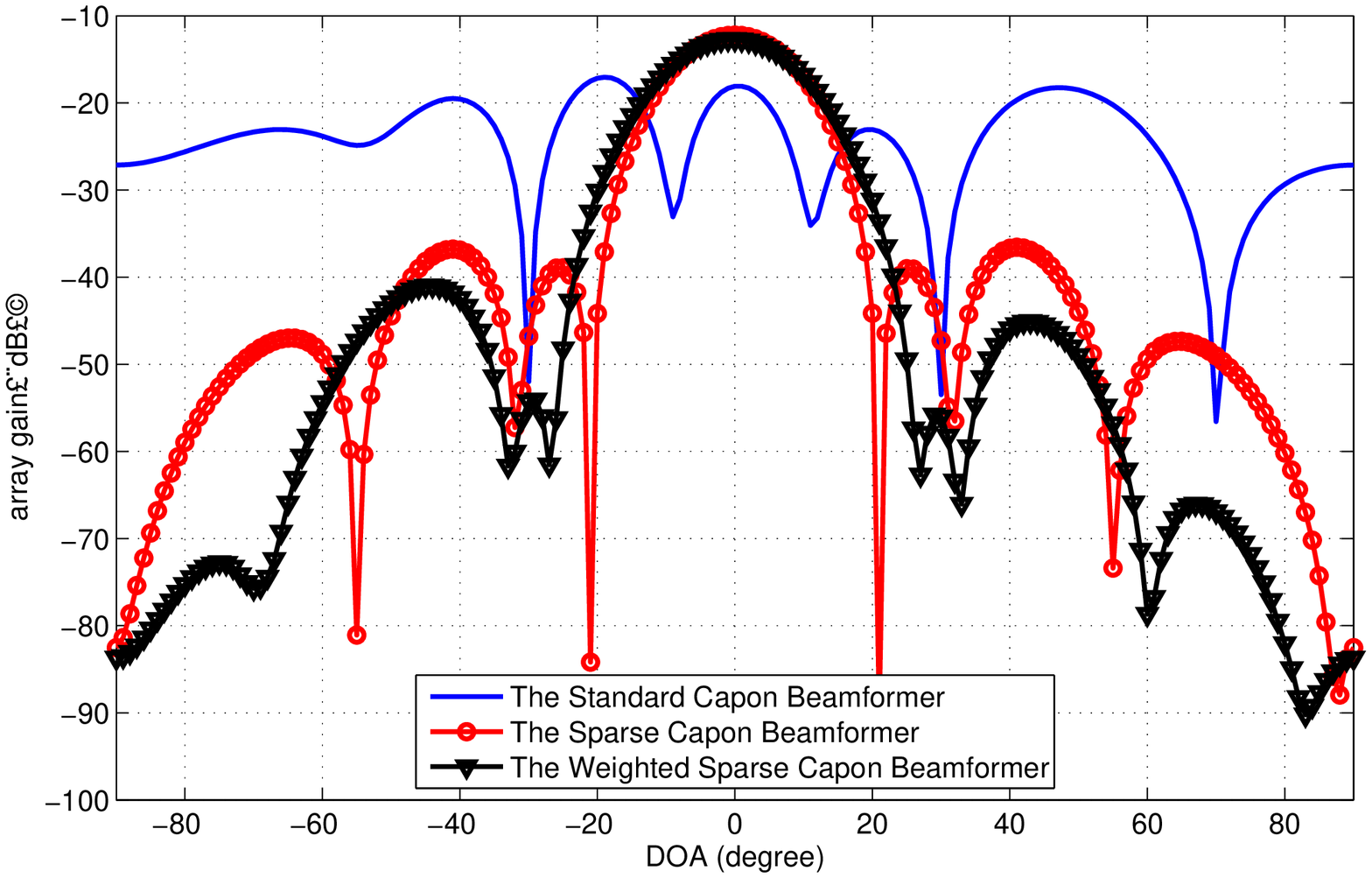}
 \caption{The beam patterns of standard Capon beamforming, sparse Capon beamforming, and weighted sparse beamforming, without any mismatch between the estimated DOA of SOI and the real one.}
 \label{figure3}
\end{figure}

\begin{figure}[!h]
 \centering
 \includegraphics[scale = 0.47]{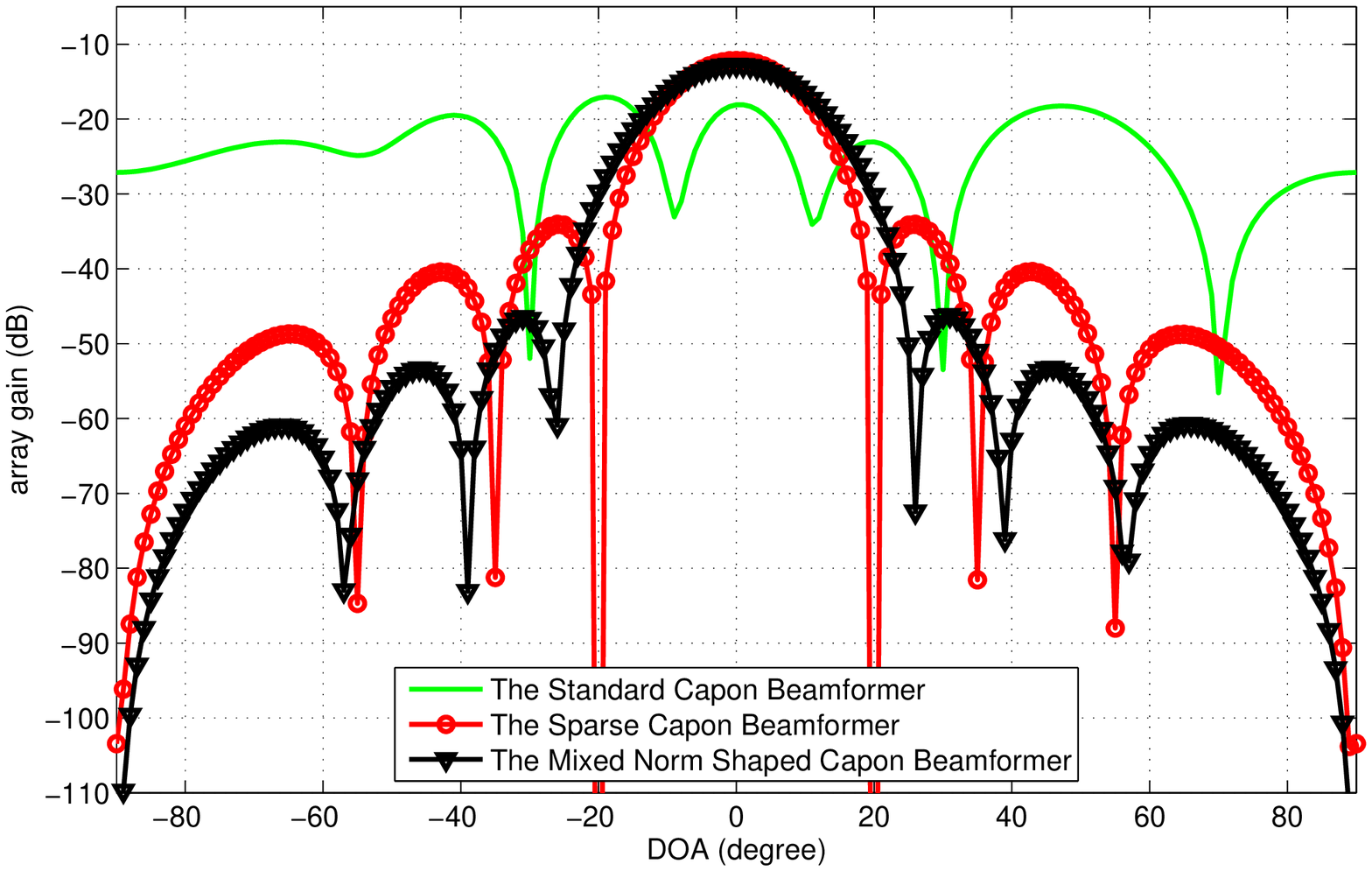}
 \caption{The beam patterns of standard Capon beamforming, sparse Capon beamforming, and mixed norm shaped Capon beamforming, without any mismatch between the estimated DOA of SOI and the real one.}
 \label{figure4}
\end{figure}

\begin{figure}[!h]
 \centering
 \includegraphics[scale = 0.47]{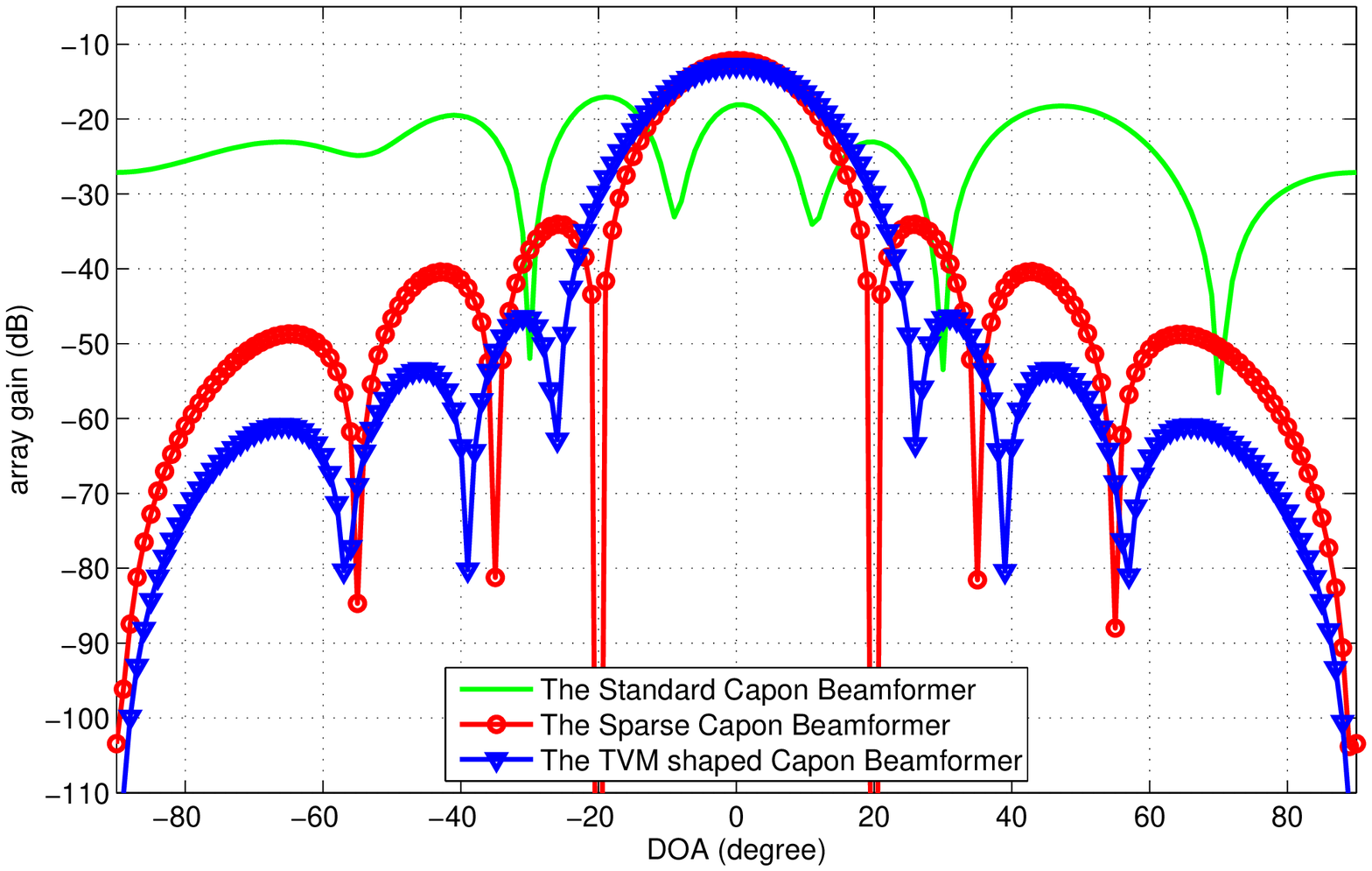}
 \caption{The beam patterns of standard Capon beamforming, sparse Capon beamforming, and TVM shaped Capon beamforming, without any mismatch between the estimated DOA of SOI and the real one.}
 \label{figure5}
\end{figure}

\begin{figure}[!h]
 \centering
 \includegraphics[scale = 0.47]{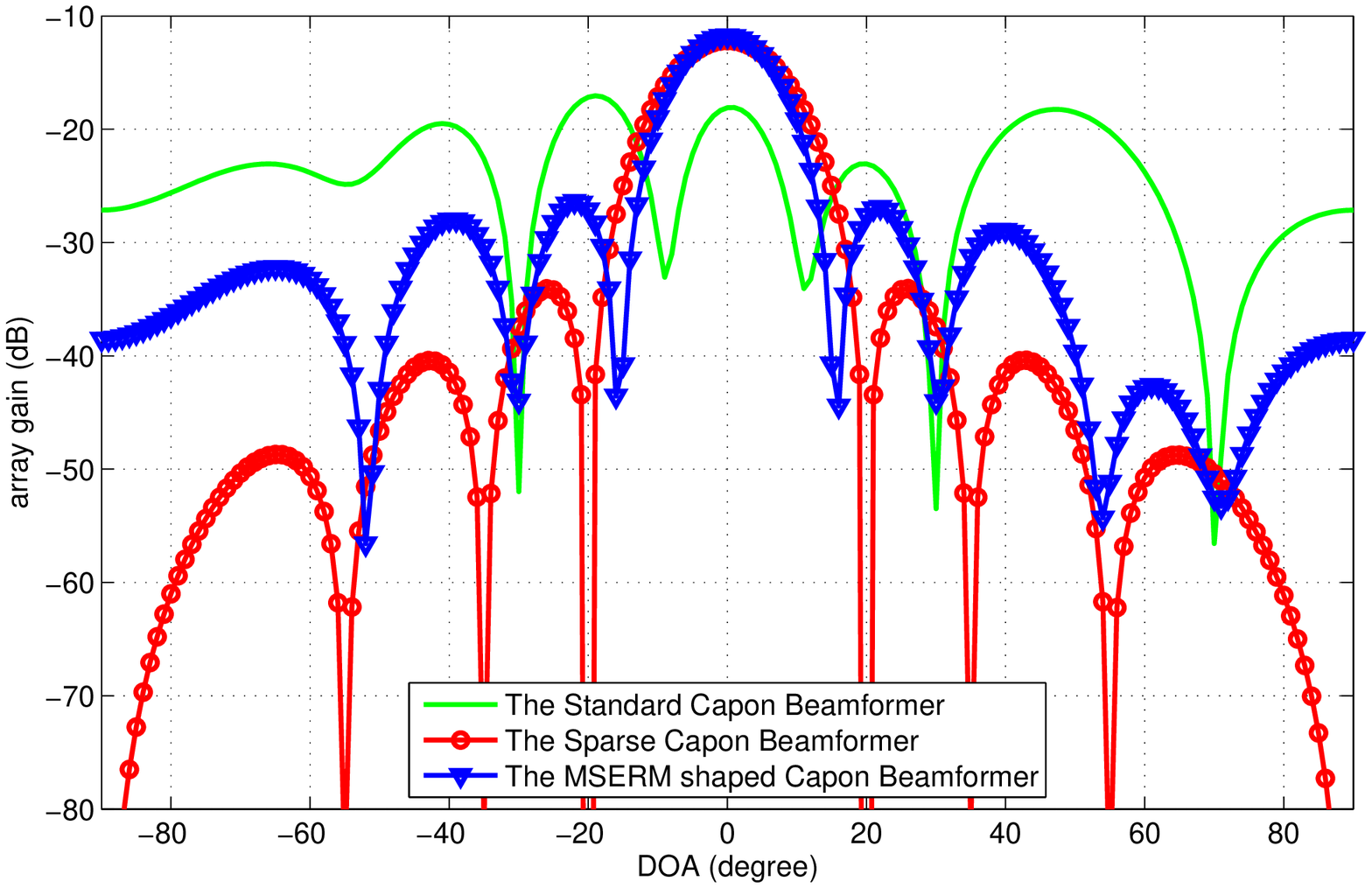}
 \caption{The beam patterns of standard Capon beamforming, sparse Capon beamforming, and MSPR shaped Capon beamforming, without any mismatch between the estimated DOA of SOI and the real one.}
 \label{figure6}
\end{figure}

When the estimated DOA of the SOI and the real one are exactly the same, i.e. $ {\theta _0} = {\alpha _0} = {0^ \circ } $, in 1000 times Monte Carlo simulations, Fig. \ref{figure3} gives the beam patterns of the Capon beamforming, sparse beamforming and weighting sparse Capon beamforming; Fig. \ref{figure4} gives the beam patterns of the Capon beamforming, the sparse Capon beamforming and the mixed norm based Capon beamforming; Fig. \ref{figure5} gives the beam patterns of the Capon beamforming, the sparse Capon beamforming and the TVM based Capon beamforming. Fig. \ref{figure6} gives the beam patterns of the Capon beamforming, the sparse Capon beamforming and the MSPR based Capon beamforming. Each beam pattern is normalized with its L2 norm, i.e. the power of the array gains is 1. Obviously it can be seen from the figures that the RCBs with pattern shaping constraints outperform the standard one. Compared with the sparse Capon beamforming, the weighted sparse Capon beamforming, the mixed norm based Capon beamforming, the TVM based Capon beamforming and the MSPR based Capon beamforming have much lower array gains in the directions of interferences ($  - {30^ \circ
} $, $  {30^ \circ } $, and $  {70^ \circ } $). They have better nulling performance for interference suppression, and  the SINR of the array output signals. In 1000 times Monte Carlo simulations, the average SINR of the beamformers are 2.3027 dB, 4.3178 dB, 5.0162 dB, 5.8119 dB, 5.8563 dB and 6.5224 dB.

\begin{figure}[!h]
 \centering
 \includegraphics[scale = 0.47]{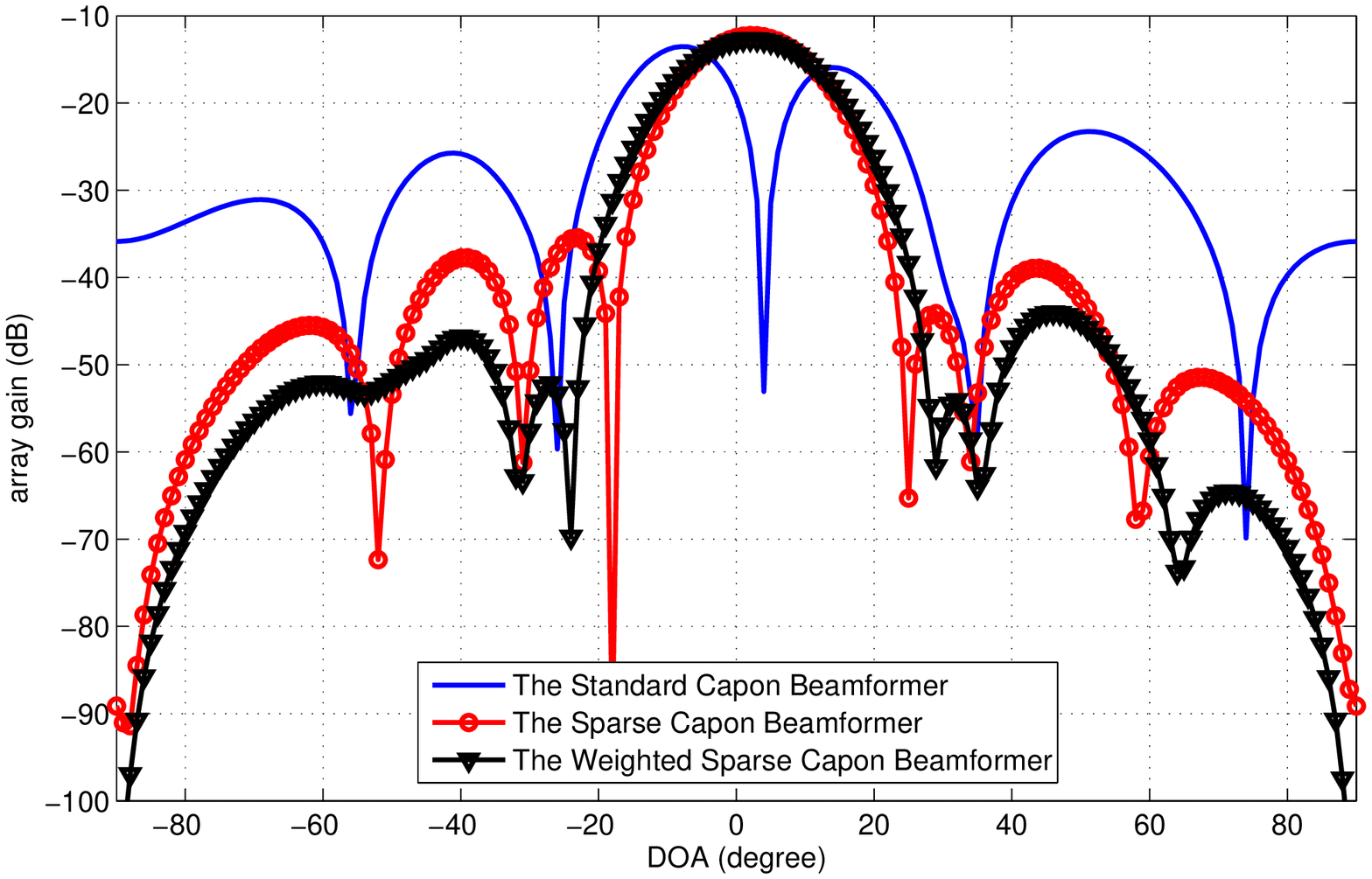}
 \caption{The beam patterns of standard Capon beamforming, sparse Capon beamforming, and weighted sparse Capon beamforming, with $ {3^ \circ } $ mismatch between the estimated DOA of SOI and the real one.}
 \label{figure7}
\end{figure}

\begin{figure}[!h]
 \centering
 \includegraphics[scale = 0.47]{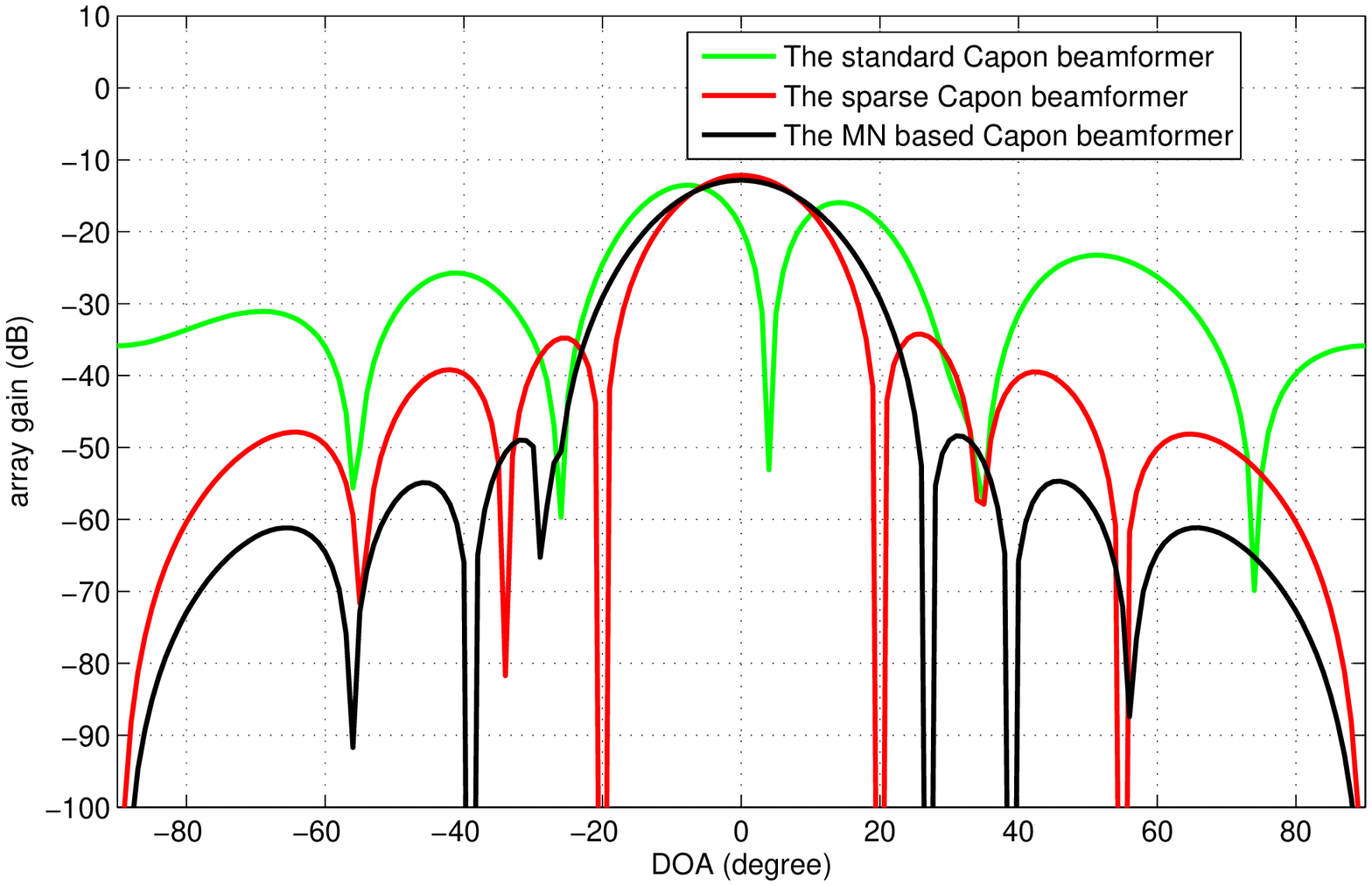}
 \caption{The beam patterns of standard Capon beamforming, sparse Capon beamforming, and mixed norm shaped Capon beamforming, with $ {3^ \circ } $ mismatch between the estimated DOA of SOI and the real one.}
 \label{figure8}
\end{figure}

\begin{figure}[!h]
 \centering
 \includegraphics[scale = 0.47]{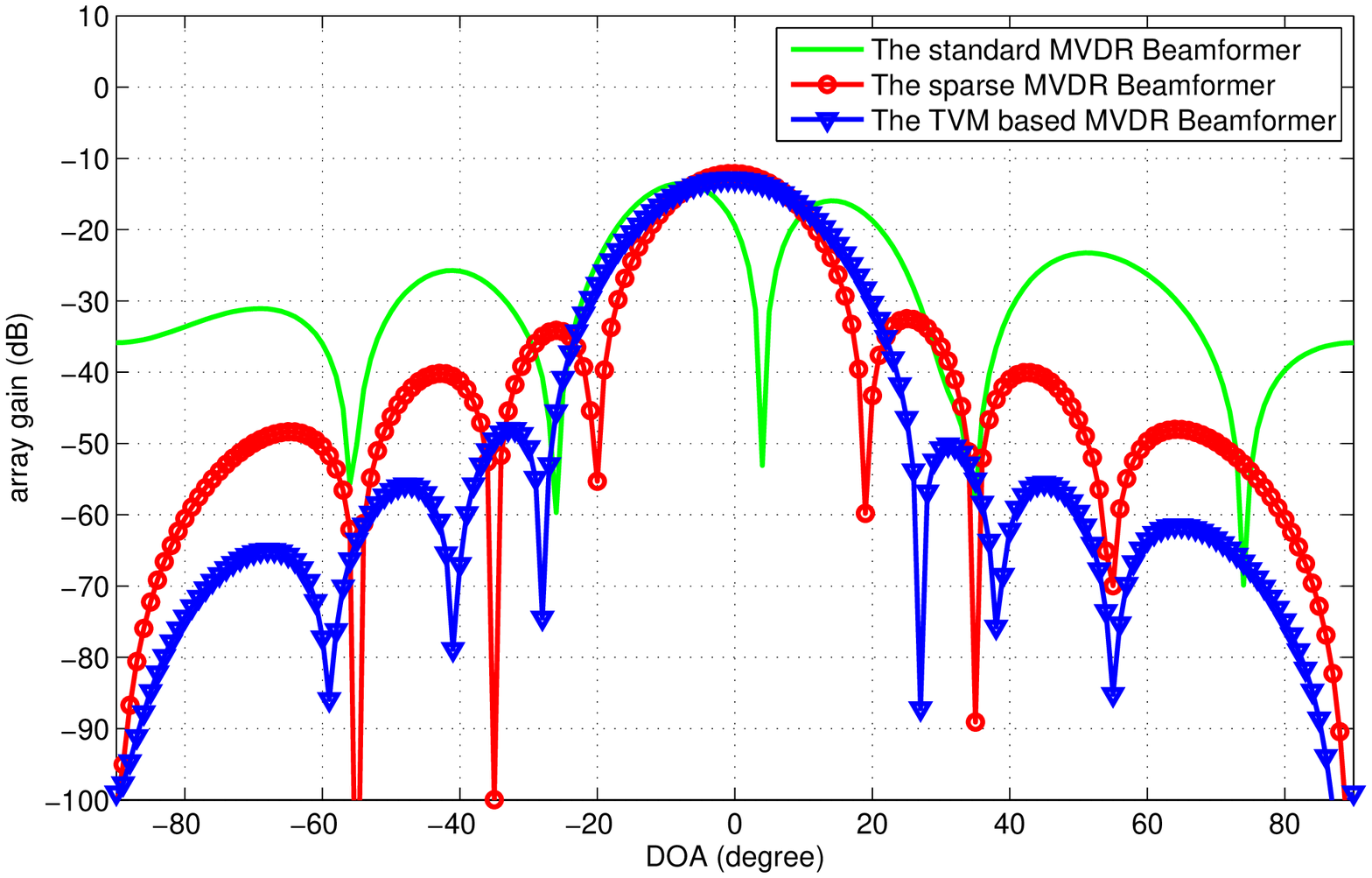}
 \caption{The beam patterns of standard Capon beamforming, sparse Capon beamforming, and TVM shaped Capon beamforming, with $ {3^ \circ } $ mismatch between the estimated DOA of SOI and the real one.}
 \label{figure9}
\end{figure}

\begin{figure}[!h]
 \centering
 \includegraphics[scale = 0.47]{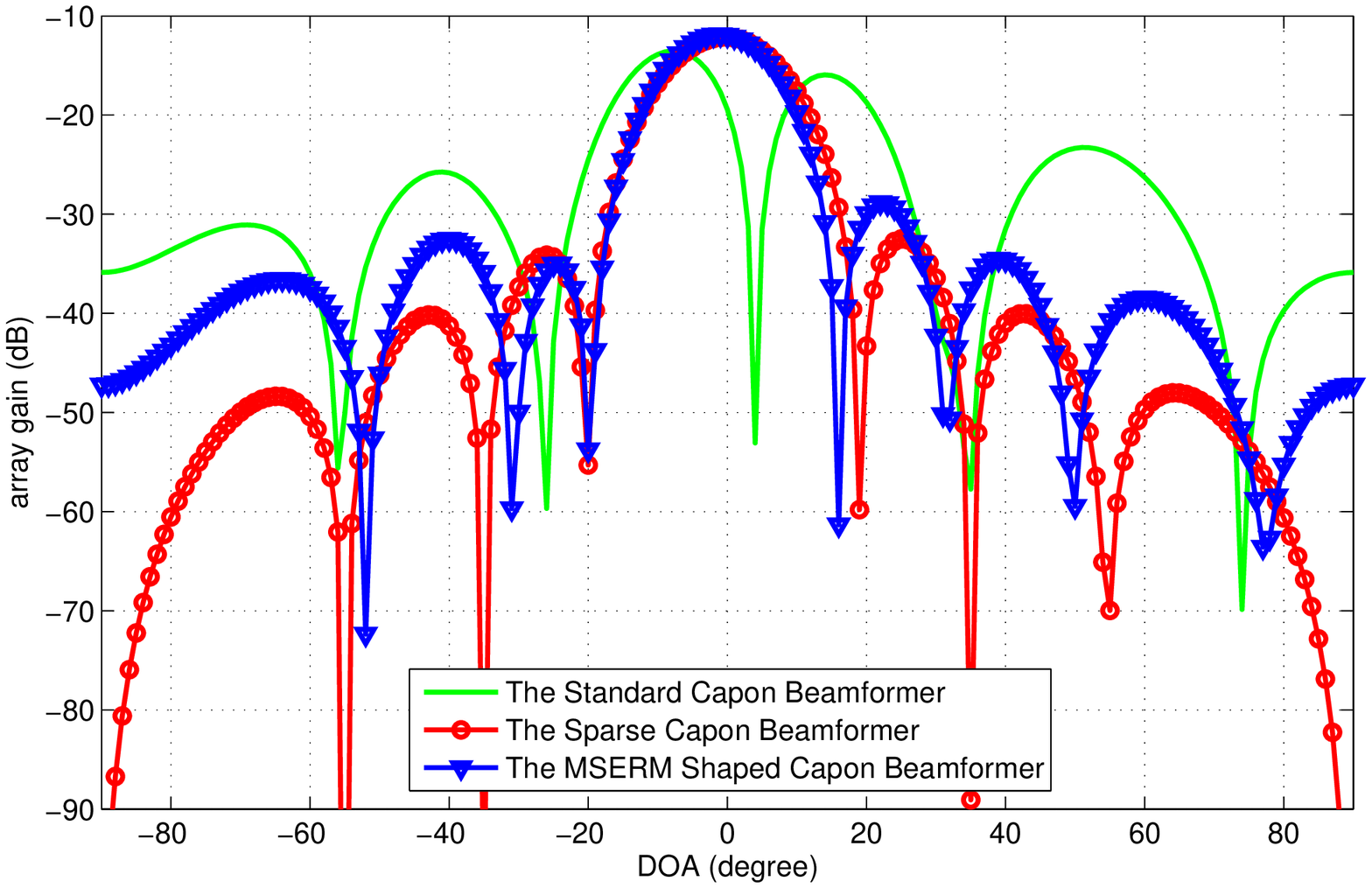}
 \caption{The beam patterns of standard Capon beamforming, sparse Capon beamforming, and MSPR shaped Capon beamforming, with   mismatch between the estimated DOA of SOI and the real one.}
 \label{figure10}
\end{figure}

In practice, the high sensitivity of angle mismatch is one of the main disadvantages of Capon beamforming. In standard Capon beamforming, when the estimated DOA of SOI is quite different from the real one, i.e. $ {\theta _0} \ne {\alpha _0} $, the distortionless constraint of the Capon beamforming is for the direction, which allows the signal from the direction $ {\alpha _0} $ is fully received. However, the actual signal is from the direction $ {\alpha _0} $, the other constraint of the Capon beamforming, i.e. the minimization of the output power of the array, would regard the SOI from direction $ {\alpha _0} $ as an interference. The minimum variance distortionless constraints would make a nulling in the direction of SOI, which would greatly decrease the SINR of the output signals. When there is a $ {3^ \circ } $  angle mismatch ($ {\theta _0} = {3^ \circ } $  and $ {\alpha _0} = {0^ \circ }$), the beam patterns are shown in Fig. \ref{figure7} and Fig. \ref{figure10}. Suppressing the sidelobe level and obtaining deeper nullings to avoid interferences, Fig. \ref{figure7}, Fig. \ref{figure8}, Fig. \ref{figure9} and Fig. \ref{figure10} show that beam pattern shaping constraint based Capon beamforming has better robustness than standard Capon beamforming. Obviously we can see that there is a nulling in the direction of SOI. However, the array gains in the real DOA of SOI have almost the same level of the estimation DOA. Similarly, compared sparse Capon beamforming, weighted sparse Capon beamforming, mixed norm based Capon beamforming, and TVM based Capon beamforming, and MSPR based Capon beamforming have lower nulling for avoiding interference. When there is angle mismatch, as denoted in table \ref{table1}, in 1000 times' Monte Carlo simulations, the values of SINR of each beamformers' output signal are: 0.0003 dB, 3.1903 dB, 3.8013 dB, 4.6836 dB, 4.8667 dB and 3.8402 dB.

In the above discussion, the SINR values show the performance of the sparse Capon beamforming is the worst of all the beam pattern shaping constraints based Capon beamforming, but it has some advantages in some other applications. Comparing the beam patterns of sparse Capon beamforming and other robust Capon beamforming methods, the beam pattern of sparse Capon beamforming has a narrow mainlobe, which is preferred in radar applications.

In addition, in the presence of 3 degrees mismatch, the SINR of MSPR constraint based Capon beamforming is less than the values of SINR of mixed norm based Capon beamforming and TVM based Capon beamforming. However, in the case of no angle mismatch, MSPR constraint based Capon beamforming has the best SINR performance. Thus we can see that MSPR constraint based Capon beamforming has best performance of suppressing sidelobe, but its robustness against the angle mismatch is inferior to the mixed norm based Capon beamforming and TVM based Capon beamforming.

To sum up, a large performance improvement can be obtained by adding the proposed beam pattern shaping constraints with respect to standard Capon beamforming. Meanwhile, Comparing the beamforming methods in this paper, each methods have its own characteristics. In certain environments, each has some unique performance advantages.

\section{Conclusion}

This paper begins with a brief introduction of the Capon beamforming research and existing problems. Then, the problems of the standard Capon beamforming are given: the high sensitivity of the DOA mismatch and the high sidelobe level. A set of novel methods are summarized to deal with these two issues. The methods used here improve the beamforming performance by adding the array gain distribution encouragement constraints. In order to get an ideal beam pattern distribution, This paper has discussed the sparse constraint, the weighted sparse constraint, the mixed norm constraint, TVM constraint, and the MSPR maximization constraint. Numerical experiments show that the performance of Capon beamforming is significantly improved by using beam shaping constraints. It preferably overcomes  the high sensitivity of angle mismatch problem and high sidelobe level. In addition, the performance of different beam-shaping constraints are compared.

In the future, the beam pattern shaping constraint can be combined with other robust Capon beamforming techniques, such as ellipsoid methods, diagonal loading methods. moreover, it is natural to generalize the beam pattern shaping constraints to the 3-D beamforming.

%
%
%

\ifCLASSOPTIONcaptionsoff
  \newpage
\fi

\begin{tablehere}
\onecolumn
\renewcommand{\arraystretch}{1.3}
\caption{The average values of SINR of the beamformers with and without $ {3^ \circ } $ DOA mismatch (dB). } \vskip0.2in
\begin{center}
\small \begin{tabular}{|c|c|c|c|c|c|c|c|c|}\hline ~ & Capon & sparse Capon & weighted sparse Capon &
mixed norm & \hspace{0.5cm}  TVM  \hspace{0.5cm} & \hspace{0.5cm} MSPR \hspace{0.5cm}
\\ \hline no DOA mismatch & 0.0281 & 0.1492 & 0.0527 & 0.2146 & 0.1012 & 0.2047
\\ \hline $ {3^ \circ } $ DOA mismatch & 0.0231 & 0.1283 & 0.0577 & 0.1642 & 0.1562 & 0.1860
 \\\hline
  \end{tabular}
\end{center}
\label{table1}
\end{tablehere}

%
%

\end{document}